\documentclass[1p,10pt]{elsarticle}

\usepackage{rotating}
\usepackage{color}
\definecolor{grey}{rgb}{0.95, 0.95, 0.95}
\usepackage[final]{listings}
\usepackage[font=footnotesize]{subfig}
\usepackage{sidecap}

\graphicspath{{figures/}}

\begin{document}
\begin{frontmatter}
%
\title{Efficient multicore-aware parallelization strategies for iterative stencil computations}
\author{Jan Treibig, Gerhard Wellein, Georg Hager}
\address{Erlangen Regional Computing Center (RRZE)\\ University Erlangen-Nuremberg\\Erlangen,Germany\\
\ead{jan.treibig@rrze.uni-erlangen.de}}
\begin{abstract} 
Stencil computations consume a major part of runtime in many
scientific simulation codes. As prototypes for this class of
algorithms we consider the iterative Jacobi and Gauss-Seidel smoothers
and aim at highly efficient parallel implementations for cache-based
multicore architectures. Temporal cache blocking is a known advanced
optimization technique, which can reduce the pressure on the memory
bus significantly. We apply and refine this optimization for a
recently presented temporal blocking strategy designed to explicitly
utilize multicore characteristics.  Especially for the case of
Gauss-Seidel smoothers we show that simultaneous multi-threading (SMT)
can yield substantial performance improvements for our optimized
algorithm.
\end{abstract}
\end{frontmatter}

%
\section{Introduction}
%
Stencil computations can be found at the core of many scientific and technical
applications based on regular lattices. For the important class of partial
differential equation (PDE) solvers they are a key performance factor. This
does not only hold for serial applications but is also true for massively
parallel large scale multigrid PDE solvers (see e.g.~\cite{bergen05}), where
the time-consuming smoothing steps are frequently composed of stencil
computations such as red-black Gauss-Seidel or Jacobi schemes.

It has been shown recently~\cite{datta08} that for state-of-the-art multicore
architectures a near to optimal stencil implementation requires elaborate
tuning, even if more complex temporal blocking techniques
are ignored. Conventional temporal blocking performs multiple updates on a
small block of the computational domain before proceeding to the next block.
Apart from having a machine dependent tuning parameter, this kind of temporal
blocking in three spatial dimensions has been found to not deliver performance
improvements \cite{kowarschik04} because it generates rather short loops
resulting in substantial performance penalties from pipeline start-up effects
and, even worse, from strongly restricted data prefetching abilities.  In
\cite{treibig09} it was shown that adapted variants of these temporal blocking
techniques, operating on whole lines instead of rectangular small blocks, show
very good results on modern architectures also in 3D. Cache oblivious
algorithms as proposed by Frigo et al.~\cite{frigo99} are hardware independent
but come at the cost of irregular block access patterns, which cause many data
TLB misses.  This was shown for a 3D lattice Boltzmann (LB) application kernel
in Ref.~\cite{zeiser08}\@. For an overview about stencil algorithm specific optimizations
we refer to \cite{datta09}.

Recently a proof of concept for a wavefront-based shared memory
parallelization scheme was first presented \cite{wellein09}. It allows implicit temporal
blocking for stencil computations in multicore environments with shared caches.
The basic idea is to run multiple wavefronts through the computational domain
at the same time but appropriately shifted in space (depending on the stencil
used)\@. Each wavefront represents an update step of the lattice and is
executed by a single thread. Binding all threads that run successive wavefronts
for the same computational domain to a single multicore chip with a shared
cache restricts data access to main memory to a load operation for the initial
wavefront and a store operation for the final wavefront; all other
(intermediate) data accesses can be satisfied from the shared cache.  The
scheme is tailored to multicore architectures with shared caches ---
which will be the major design principle for most standard architectures in the
years to come --- and does not exhibit the drawback of very short loop lengths.

This paper introduces a more generic and flexible implementation of the
wavefront technique with an improved spatial blocking scheme and highly
efficient synchronization primitives. In addition to the application on Jacobi,
already shown in \cite{wellein09}, we extend the method to the lexicographic
Gauss-Seidel smoother, whose data dependiencies require substantial
modifications in thread scheduling \cite{hager03}. Moreover we show how our technique
can leverage SMT threads on Intel processors increasing the utilization of the
floating point units.


All modern x86-based compute nodes employ scalable ccNUMA architectures and
thus we focus on single socket results only. The parallelization strategies
used in this report are known to scale well on these architectures.

%
\section{Experimental Testbed}
\label{sec:testbed}
\begin{figure}[htbp]\centering
    \subfloat[Core 2 Quad]{\includegraphics*[width=0.4\linewidth]{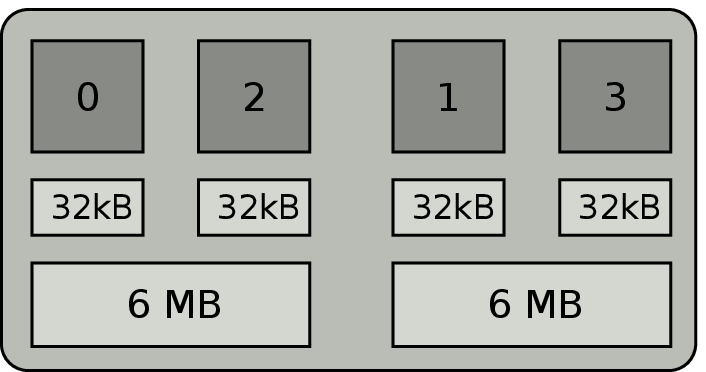}}\qquad
    \subfloat[Nehalem EP Westmere]{\includegraphics*[width=0.5\linewidth]{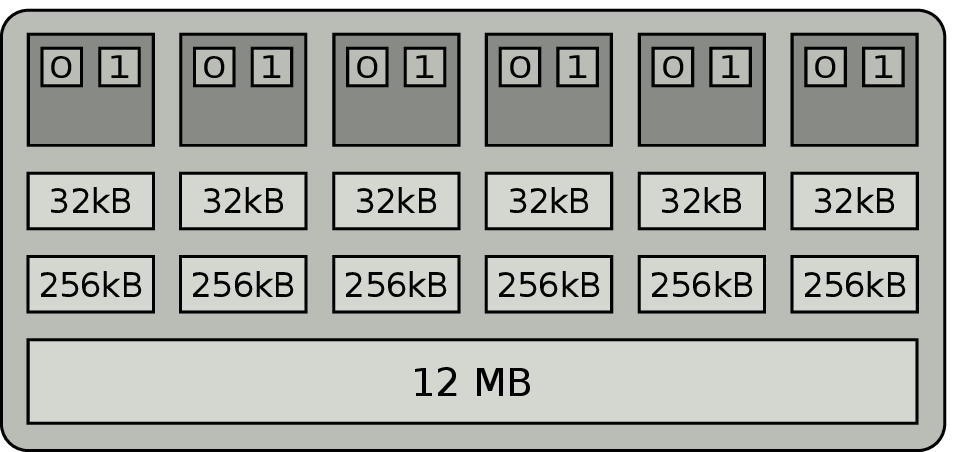}}
    \caption{Cache and thread topology of Intel Core 2 Quad and Nehalem EP Westmere processors.}
    \label{fig:topology}
\end{figure}

%
A wide selection of modern x86-based multi-core processors (cf.
Tab.~\ref{tab:arch}) has been chosen to try different variants of the wavefront
parallelization strategy and to demonstrate its performance potentials. All of
these chips feature a large outer level cache, which is shared by two (Intel
Harpertown), four (Intel Nehalem EP), six (Intel Westmere EP, AMD Istanbul) or
eight cores (Intel Nehalem EX). The maximum number of cores sharing an outer
level L2/L3 cache will be denoted as ``L2/L3 group'' in the following. The
Harpertown processor (implementing the Core 2 architecture) is usually
considered as a quad-core chip since four cores are put in a single package
(see Fig.~\ref{fig:topology}~(a)). However, it is built up from two independent
L2 groups without a shared L3 cache and thus we will consider it as two
independent dual-core processors, i.e. two L2 groups. The Nehalem EP is Intel's
first quad-core chip featuring a shared L3 cache for all four cores (L3 group).
In addition a complete redesign of the memory subsystem allowed for a
substantial increase in main memory bandwidth at the cost of introducing a
ccNUMA architecture for multi-socket servers. The follow-on processor
(Westmere) reflects a shrink in transistor size, which allows to  increase both
the number of cores as well as the L3 cache size by 50\,\%
(Fig.~\ref{fig:topology}~(b))\@. Intel also reintroduced simultaneous
multithreading (SMT) with the Nehalem architecture, a hardware optimization to
improve utilization of execution units. Each core supports two SMT threads.
AMD's competitor to Intel Nehalem is the Istanbul processor design, which comes
as a six core L3 group and is based on a ccNUMA architecture on the
multi-socket node level as well. The 8-core Intel Nehalem EX processor is not
mainly targeted at HPC clusters but at large mission-critical servers. Since it
already implements a substantially improved cache architecture and can easily
be manipulated on the main memory bandwidth side we have included it in this
report to simulate future architectural developments. A comprehensive summary
of the most important processor features is presented in Tab.~\ref{tab:arch}.
As the two iterative schemes considered in this paper are known to be memory
bandwidth intensive we have reported in Tab.~\ref{tab:arch} the maximum
attainable main memory bandwidth as measured by the STREAM triad benchmark
\cite{stream} with and without non temporal stores. In the latter case the full
bus traffic including the write allocate transfer for the store stream is reported. For a
more detailed analysis of the memory and cache hierarchy see~\cite{Jan09}.
Please note, that our Intel Nehalem EX test system was equipped with only half
of the possible number of memory cards, reducing the bandwidth accordingly. The
ability of pinning a selected team of threads to a single cache group and
determining the cache group topologies of a multi-core processor is vital for
the parallelization approach described in this report. In this context we use
the open-source tool likwid~\cite{likwid}, which has been developed in our
group.
\begin{table}[p]
    \caption{Test machine specifications. The cacheline size is 64 bytes 
	for all processors and cache levels. Note that the Nehalem EX system was only equipped
        with half the required memory boards, reducing main memory bandwidth by 50\,\%. The STREAM
        results were obtained with the triad variant.}
    \label{tab:arch}
    \centering\footnotesize
        \begin{sideways}\renewcommand{\arraystretch}{1.5}
	\begin{tabular}{lccccc}
	    \hline
	    Microarchitecture           &Intel Core 2      &Intel Nehalem EP  &Intel Westmere   &Intel Nehalem EX   &AMD Istanbul      \\
	    Model                       &Xeon X5482        &Xeon X5550        &Xeon X5670       &Xeon X7560         &Opteron 2435  \\
	    \hline                                                                                                                      
	    Clock [GHz]                 &3.2               &2.66              &2.93             &2.26               &2.6               \\
        Cores per socket            &4                 &4                 &6                &8                  &6                 \\
        SMT threads per core        &N/A               &2                 &2                &2                  &N/A               \\
	    \hline                                                                                                                     
	    L1 Cache                    &32 kB             &32 kB             &32 kB            &32 kB              &64 kB             \\
	    Associativity               &8                 &8                 &8                &8                  &2                 \\
	    \hline                                                                                                                     
	    L2 Cache                    &2x6 MB (shared)   &4x256 KB          &6x256 KB         &8x256 KB           &6x512 KB          \\
	    Associativity               &24                &8                 &8                &8                  &16                \\
	    \hline                                                                                                                     
	    L3 Cache (shared)           &-                 &8 MB              &12 MB            &24 MB              &6 MB              \\
	    Associativity               &-                 &16                &16               &24                 &48                \\
	    \hline                                                                                                                     
            Bandwidths [GB/s]:          &                  &                  &                 &                   &       \\\hline 
        Theoretical socket BW           &12.8              &32.0              &32.0             &17.1               &17.1              \\
	    STREAM 1 thread             &4.6               &11.9              &11.0             &5.3                &7.2               \\
        STREAM socket NT/noNT          &4.8/5.6           &18.5/23.7         &21.0/23.6        &9.1/13.6          &9.8/11.4           \\
	    \hline
	\end{tabular}
      \end{sideways}
\end{table}
%
\section{Iterative schemes and baseline performance}
\label{sec:code_analysis}
%
Iterative schemes based on regular stencil computations in three spatial
dimensions are used in many numerical applications, e.g. linear solvers or
multi-grid methods. As prototypes for this class of algorithm we have chosen
the well-known Jacobi method for a Poisson problem and the Gauss-Seidel method
for a Laplace problem. For reasonably large data sets those methods are known
to be data-intensive and the attainable main memory bandwidth imposes an upper
limit for performance, which can be rather accurately modeled
\cite{datta08,wellein09}. In this section we will first briefly introduce both
schemes, determine an upper performance limit (via main memory bandwidth as
given in Tab.~\ref{tab:arch}), and introduce in case of Gauss-Seidel a simple code
optimization to achieve the expected performance number. Those optimal
measurements will be the baseline to compare with for the multi-core aware
parallelization approach.

\begin{figure}[tb]\centering
    \includegraphics*[width=0.8\linewidth]{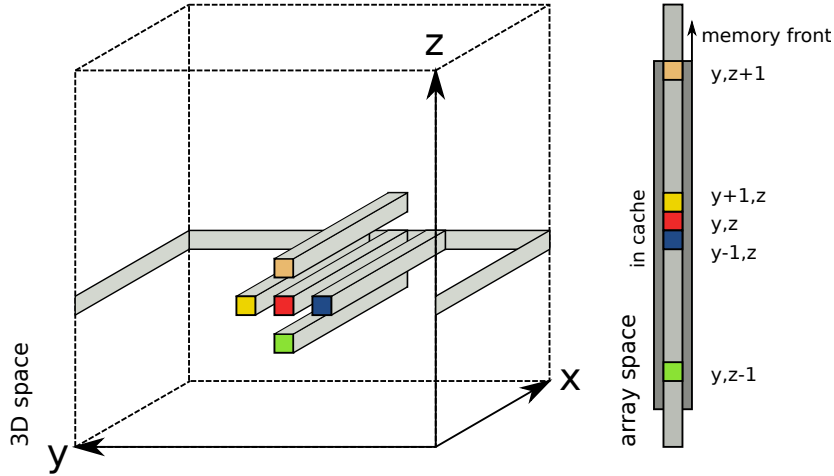}
    \caption{Stencil structure and corresponding mapping in memory space.}
    \label{fig:stencil_mapping}
\end{figure}

\begin{figure}[tb]
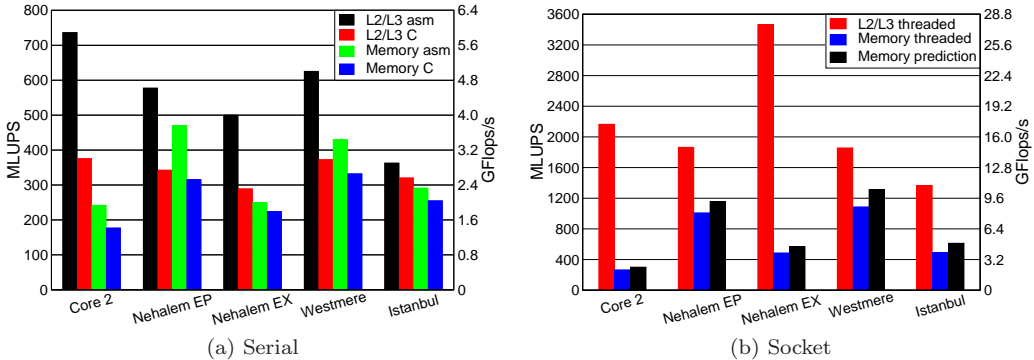
\centering
	\subfloat[Serial]{\includegraphics*[width=0.48\linewidth]{serial-jacobi.eps}}\hfill
	\subfloat[Socket]{\includegraphics*[width=0.49\linewidth]{threaded-jacobi.eps}}
        \caption{Jacobi baseline for C language (C) and assembly (asm) kernels,
        comparing memory and outer-level cache performance.  Domain sizes were
        chosen as 100x50x50 (4~MB data set fitting in the outermost cache
        level) and 400x200x200 (256~MB data set to be loaded from main memory),
        respectively. For all machines all physical cores of a socket were
        utilized.}
    \label{fig:base_jacobi}
\end{figure}

The Jacobi scheme in three spatial dimensions can basically be formulated as 
follows (we use C notation):
\begin{lstlisting}
for(iter=0; iter<iterEnd;iter++) {
 for(k=0; k<Nk; k++) {
  for(j=0; j<Nj; j++) {
   for(i=0; i<Ni; i++) {
    dst[k][j][i] = a * src[k][j][i] + b * (
                   src[k][j][i-1] + src[k][j][i+1] +
                   src[k][j-1][i] + src[k][j+1][i] + 
                   src[k-1][j][i] + src[k+1][j][i]);
} } } }
\end{lstlisting}
%
Fig.~\ref{fig:stencil_mapping} shows the basic update scheme of this kernel.
The domain is decomposed into lines ($y$ dimension) and planes ($z$ dimension).
The computational kernel updates one line at a time, and the seven point
stencil in 3D is mapped to a memory access pattern with five streams, only one
of which has  to be loaded from memory if three planes fit in the outermost
cache level (see Fig.~\ref{fig:stencil_mapping} right). The store on the
\verb+dst+ array generates an additional data stream if implemented using non temporal stores. We implement an
optimized version  of the innermost loop (\emph{line update kernel}).  This
subroutine is used for all parallel variants which only modify the processing
order of the outer loop nests.  This ensures results comparable to the
optimized variants in~\cite{datta08}.
Fig.~\ref{fig:base_jacobi} (a) shows the serial baseline performance of
the Jacobi kernel against a straightforward C implementation for
the L2/L3 cache group and main memory. The C code was compiled with Intel
compiler icc 11.0 and standard optimization flags 
(\verb+-O3  -xW -align -fno-alias+). The
same executable was used for all machines. Note that the C code could
probably reach better performance by applying
code transformations or including pragmas/intrinsics.  
Still this comparison shows the
effectiveness of the optimization techniques independent of their
implementation level. For the memory domain a variant using non-temporal
(streaming) stores was considered. As expected the highly clocked but 
bandwidth-starved Harpertown processor shows the largest drop between in-cache and main
memory performance.  The in-cache performance for the Nehalem variants is
directly correlated with their clock speed. On Nehalem EP and Westmere the
small drop between in-cache and main memory performance shows that the serial
Jacobi method is not primarily memory bandwidth limited on this machine. The
Istanbul, despite its theoretically similar capabilities, shows a low performance,
and there is no significant difference between optimized and C or in-cache and
memory performance. The combination of
exclusive caches and large cache latency overhead causes that a major part of
the runtime has to be spent transferring within the cache hierarchy (cf. ~\cite{Jan09}).
Therefore also the applied optimizations
do not show a larger effect, and the actual processing of cachelines is only a
small part of total runtime. In opposite all Intel processors show a high
cache efficiency for these bandwidth demanding data access patterns.

Turning to multi-threaded execution on a single socket one can assume that main
memory bandwidth is saturated. Following above discussion, the minimum data
transfer between main memory and cache hierarchy for a single cell update is
one load and one store. As a performance measure we use the lattice site updates per second (LUP/s)
metric. In this case a simple model for the maximum performance in terms of LUP
can be set up (for double precision):
\begin{equation}
    P_0 =  \frac{M_S}{16\;\mathrm{bytes}}[\mbox{LUP/s}]
    \label{eq:iter_limit}
\end{equation}
The computer`s attainable main memory bandwidth ($M_S$) can be measured e.g.
with the threaded STREAM triad benchmark. This simple approach is known to
provide a good upper performance limit for memory bandwidth limited situations.

We estimate the potential performance gain for temporal blocking by
benchmarking the saturated L2/L3 cache group performance with a dataset that
fits in the outer cache level. The larger the difference between in-cache and
main memory performance, the higher the expected performance improvement by
temporal blocking. For a more elaborate prediction based on a diagnostic
performance model, cf.~\cite{georg10}.  Figure~\ref{fig:base_jacobi} (b) shows
the saturated L2/L3 cache group and saturated main memory performance, together
with the limit predicted by (\ref{eq:iter_limit}). Nehalem EP and Westmere show
a more balanced performance between main memory and cache.  Hence they are
expected to benefit less from our optimizations. The measurements reveal that
while Westmere clocks higher, the uncore (L3 cache and memory controller) has
the same clock speed as Nehalem EP and therefore reaches similar in-cache
performance, despite its two additional cores. Nehalem EX introduces a novel segmented
L3 cache which shows a near to perfect bandwidth scaleup with the
number of cores. The memory results are well in line with the performance limits
predicted based on the STREAM triad sustained bandwidth as listed in
Tab.~\ref{tab:arch}.

\begin{figure}[tb]
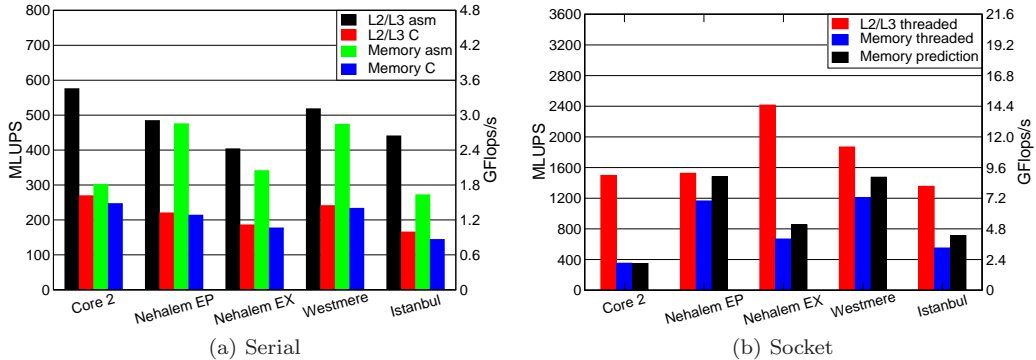
\centering
	\subfloat[Serial]{\includegraphics*[width=0.48\linewidth]{serial-gs.eps}}\hfill
	\subfloat[Socket]{\includegraphics*[width=0.49\linewidth]{threaded-gs.eps}}
    \caption{Gauss-Seidel baseline performance. Note that the C implementation 
    does not employ the dependency optimization described in the text. Same data sets as in Fig.~\ref{fig:base_jacobi}}.
    \label{fig:base_gs}
\end{figure}

%
The Gauss-Seidel method as opposed to Jacobi performs an in-place update.
It can be formulated as follows (using C notation):
\begin{lstlisting}
for(iter=0; iter<iterEnd;iter++) {
 for(k=0; k<Nk; k++) {
  for(j=0; j<Nj; j++) {
   for(i=0; i<Ni; i++) {
    src[k][j][i] =  b * ( src[k][j][i-1] + src[k][j][i+1] +
                   src[k][j-1][i] + src[k][j+1][i] + 
                   src[k-1][j][i] + src[k+1][j][i]);
} } } }
\end{lstlisting}
Gauss-Seidel looks similar to Jacobi at first glance  in terms of stencil
structure and data transfer volumes. A difference apparent at once is that SIMD
vectorization is ruled out because of the recursive structure on the central
line. The in-place update prevents optimal pipelining and the use of non-temporal stores as well.  Thus
Gauss-Seidel performance is inferior to Jacobi despite comparable data transfer volumes and less
computations (Fig.~\ref{fig:base_jacobi} and
~\ref{fig:base_gs}). Additionally there is no substantial drop between in-cache and memory
performance. This is especially remarkable on the Core 2 processor and clearly
indicates that pipelining problems limit the achievable performance. To
overcome these problems, our optimized kernel interleaves two updates in order
to break up register dependencies and partially hide the recursion. The
optimized version implements the following adjusted update scheme:
\begin{lstlisting}
for(iter=0; iter<iterEnd;iter++) {
 for(k=0; k<Nk; k++) {
  for(j=0; j<Nj; j++) {
   tmp1 = src[k][j][2] +
          src[k][j-1][1] + src[k][j+1][1] +
          src[k-1][j][1] + src[k+1][j][1];
   for(i=1; i<Ni-1; i++) {
    tmp2 = src[k][j][i+1] +
           src[k][j-1][i] + src[k][j+1][i] + 
           src[k-1][j][i] + src[k+1][j][i];

    src[k][j][i] =  b * ( src[k][j][i-1] + tmp1);
    tmp1 = tmp2;
} } } }
\end{lstlisting}
%
%
%
Figure~\ref{fig:base_gs} (a) shows the serial Gauss-Seidel results. A large
part of the speedup between the optimized assembly implementation as compared to
the C version can be attributed to this Gauss-Seidel specific code reordering.
The small drop from in-cache to memory performance indicates that the
sequential Gauss-Seidel on Nehalem EP and Westmere is not memory bandwidth
limited anymore. Only on the bandwidth starved Harpertown there is still a
significant drop between in-cache and main memory domain. Istanbul shows a much
more competitive performance for the optimized code. The data transfers are
still inefficient, but  the optimizations can show their effect and the impact
of the inefficient caches is smaller because now the L1 runtime part is much
larger. Westmere benefits from its two additional cores compared to Nehalem EP
indicating that the L3 bandwidth is not fully saturated by four threads. 

\begin{figure}[tb]
\centering
\subfloat[Pipeline parallelization]{\includegraphics*[width=0.4\linewidth]{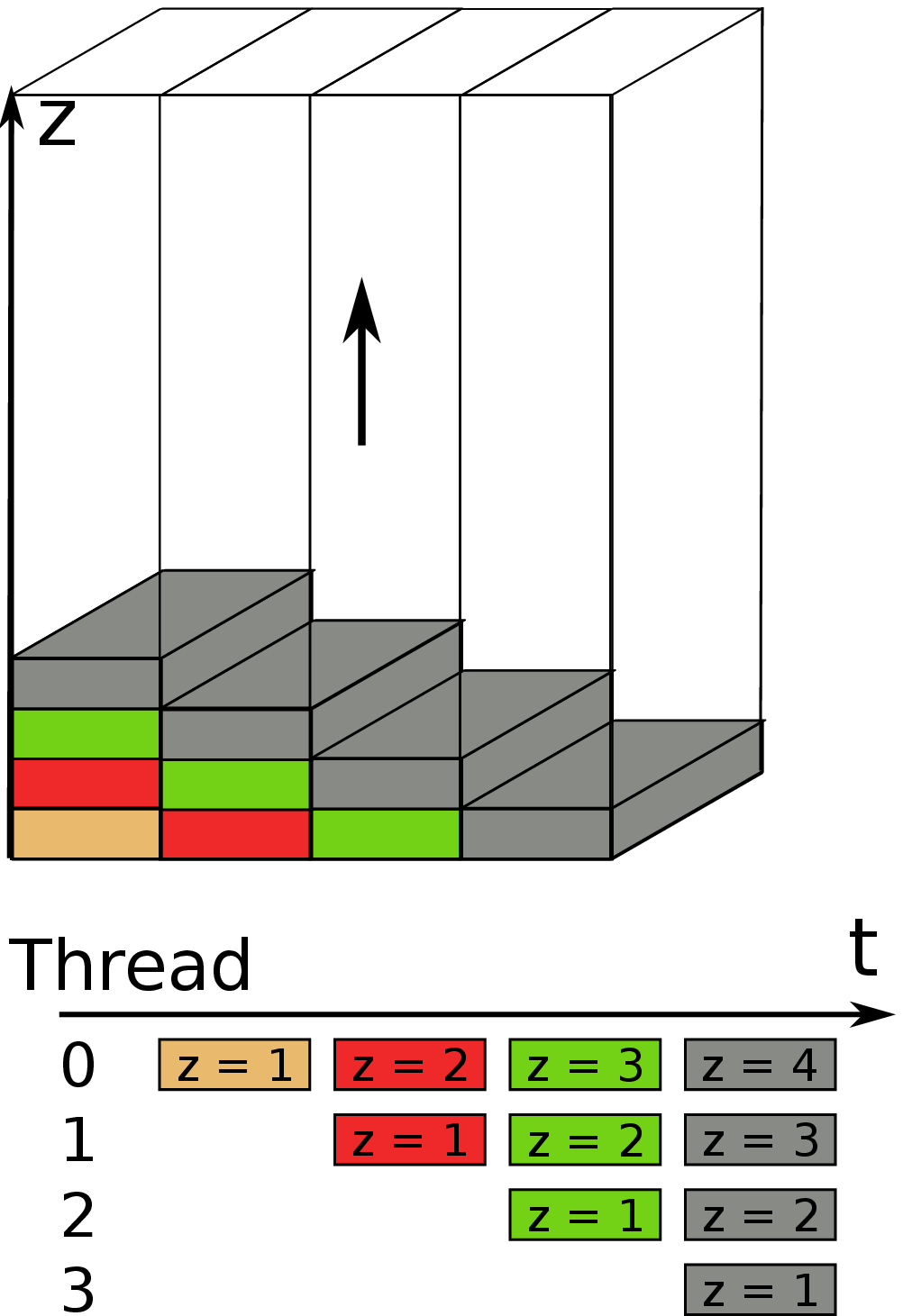}}\qquad
\subfloat[Wavefront approach]{\includegraphics*[width=0.5\linewidth]{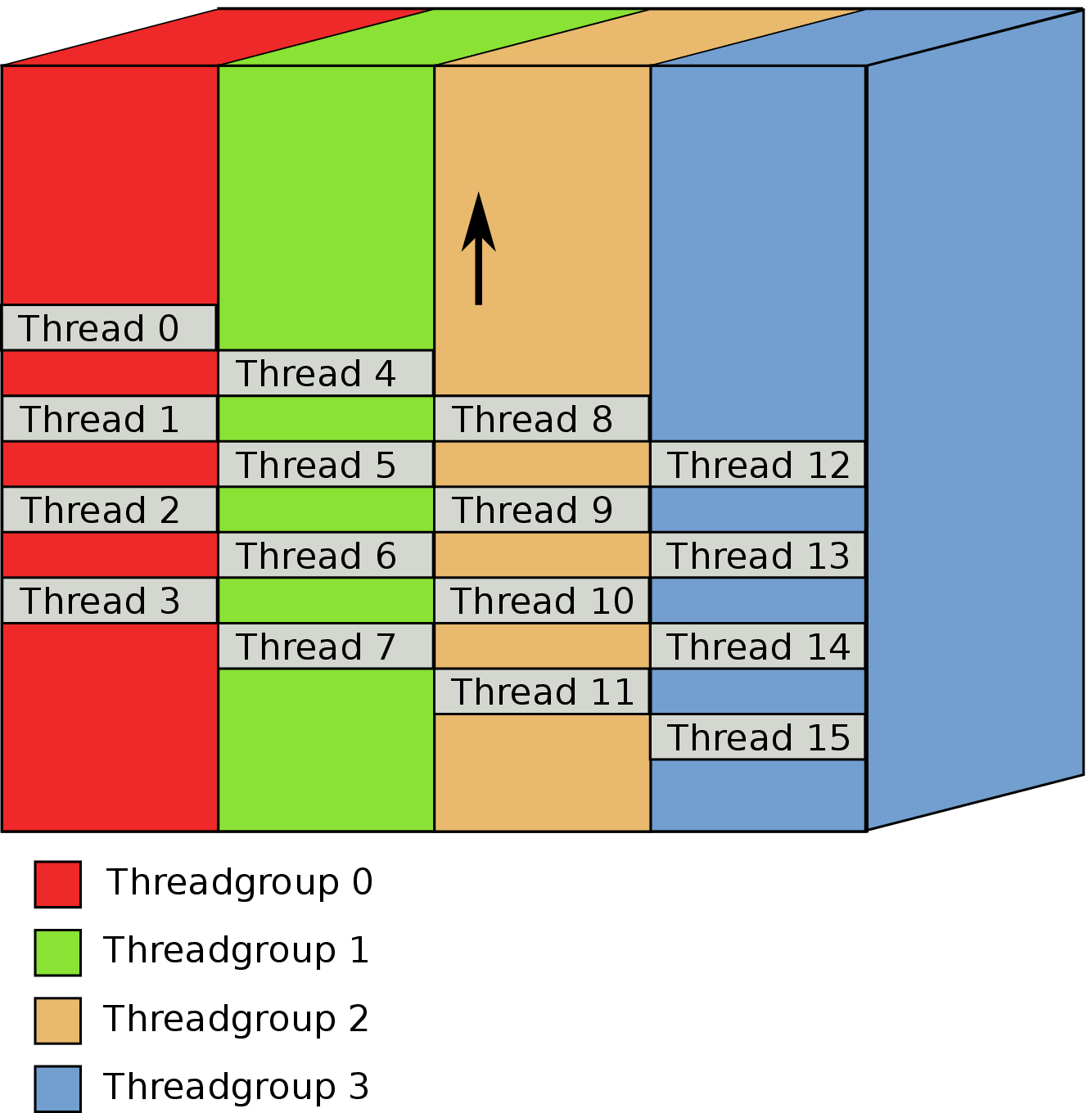}}
\caption{Parallelization of the Gauss-Seidel algorithm by pipelied parallel execution (a)
  and the wavefront approach (b)}
\label{fig:pipeline_gs}
\end{figure}

A further problem connected to the recursive nature of Gauss-Seidel is that a
straightforward parallelization based on domain decomposition cannot be
employed. A common solution is to use the Red-Black Gauss-Seidel method
instead, which can be easily parallelized. We chose another possibility to
parallelize the standard lexicographic Gauss-Seidel method based on a pipeline
parallel approach (see Fig.~\ref{fig:pipeline_gs} (a)), which retains the same
algorithm as a sequential Gauss-Seidel. Each thread operates
on a sub-block. Plane updates of threads are shifted in time to retain the
correct update order. The threaded socket results for Gauss-Seidel are
illustrated in Fig.~\ref{fig:base_gs}.
%
Compared to the Jacobi solver, the parallel Gauss-Seidel algorithm is still
limited by the loop-carried dependencies in the kernel, which leads to a
smaller performance difference between in-cache and memory situations for all
but the Westmere and Istanbul processors. Westmere can still hit the bandwidth
limitations due to its larger core count, while Istanbul is known to be
restrained by the large overheads for cache line transfers~\cite{Jan09}, making
the inefficient pipelining less dominant.

As mentioned above, non-temporal stores cannot be applied. We therefore use the
STREAM triad measurements without non-temporal stores (Tab. \ref{tab:arch}) in
the performance model (\ref{eq:iter_limit}) for Gauss-Seidel.

%
%
%
\section{Temporal Blocking through Multi-Core Aware Wavefront Parallelization}
\label{sec:optimization}
%
\begin{figure}[htbp]
\centering
\subfloat[]{\includegraphics*[width=0.4\linewidth]{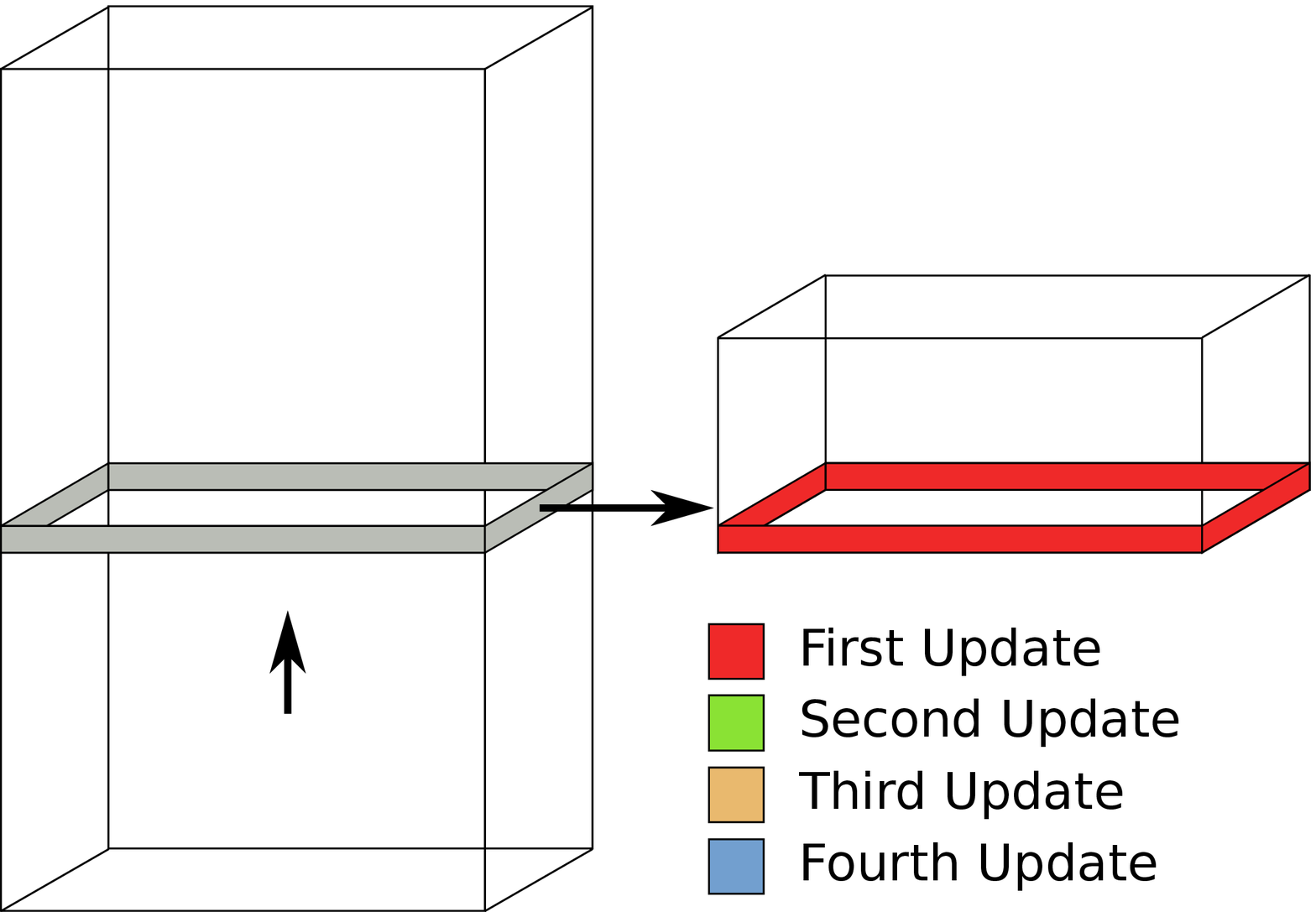}}\qquad
\subfloat[]{\includegraphics*[width=0.4\linewidth]{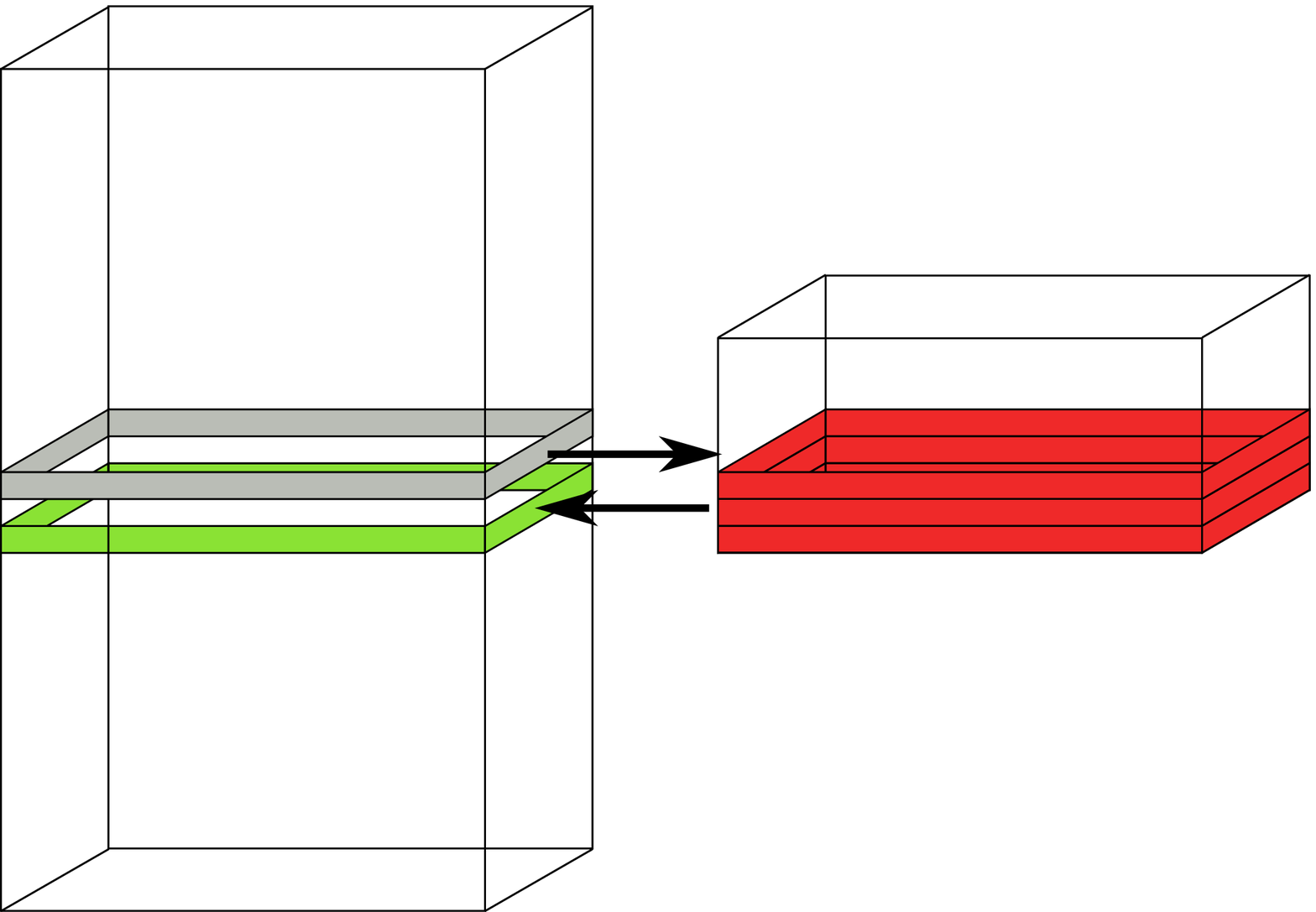}}\\
\subfloat[]{\includegraphics*[width=0.4\linewidth]{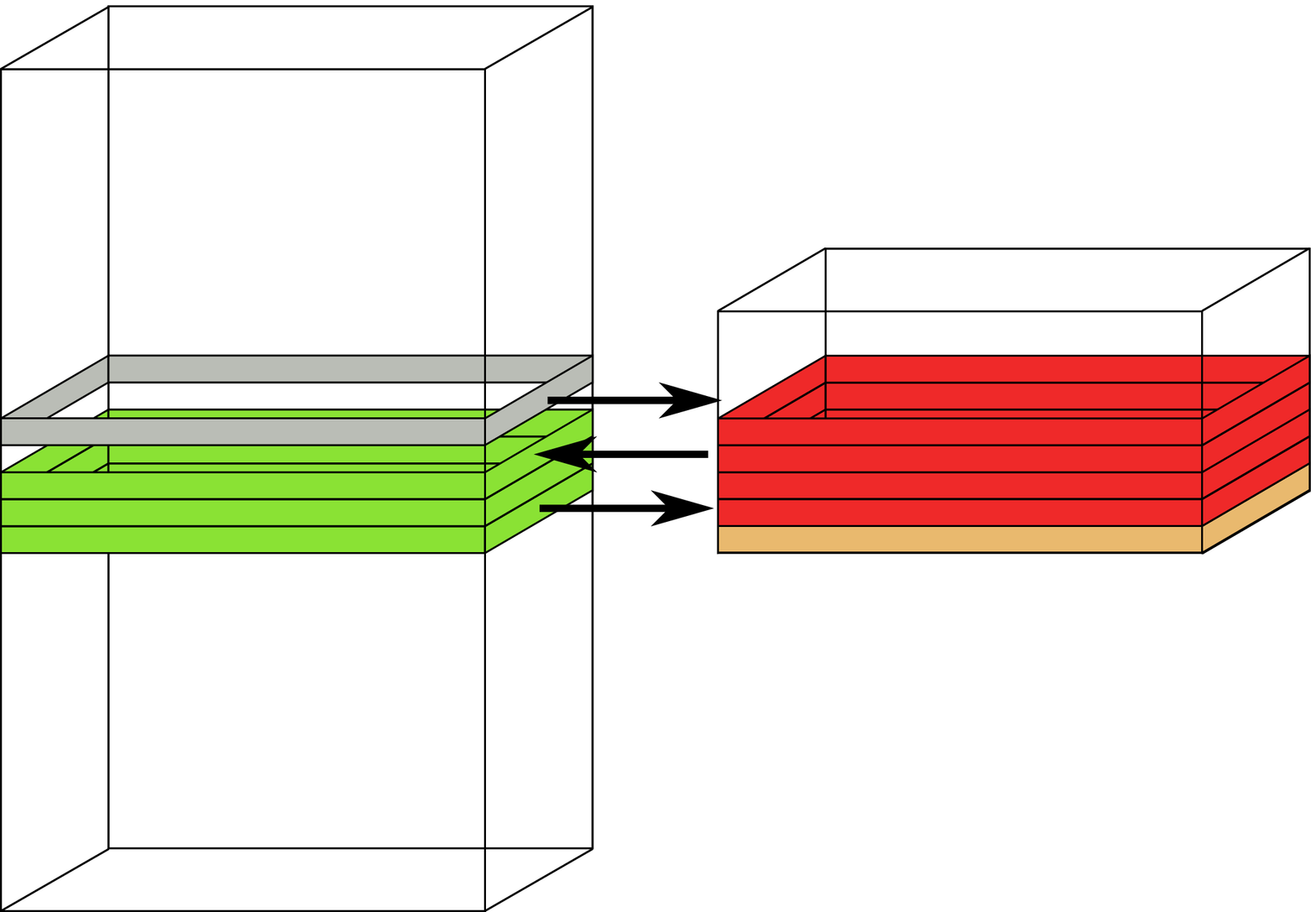}}\qquad
\subfloat[]{\includegraphics*[width=0.4\linewidth]{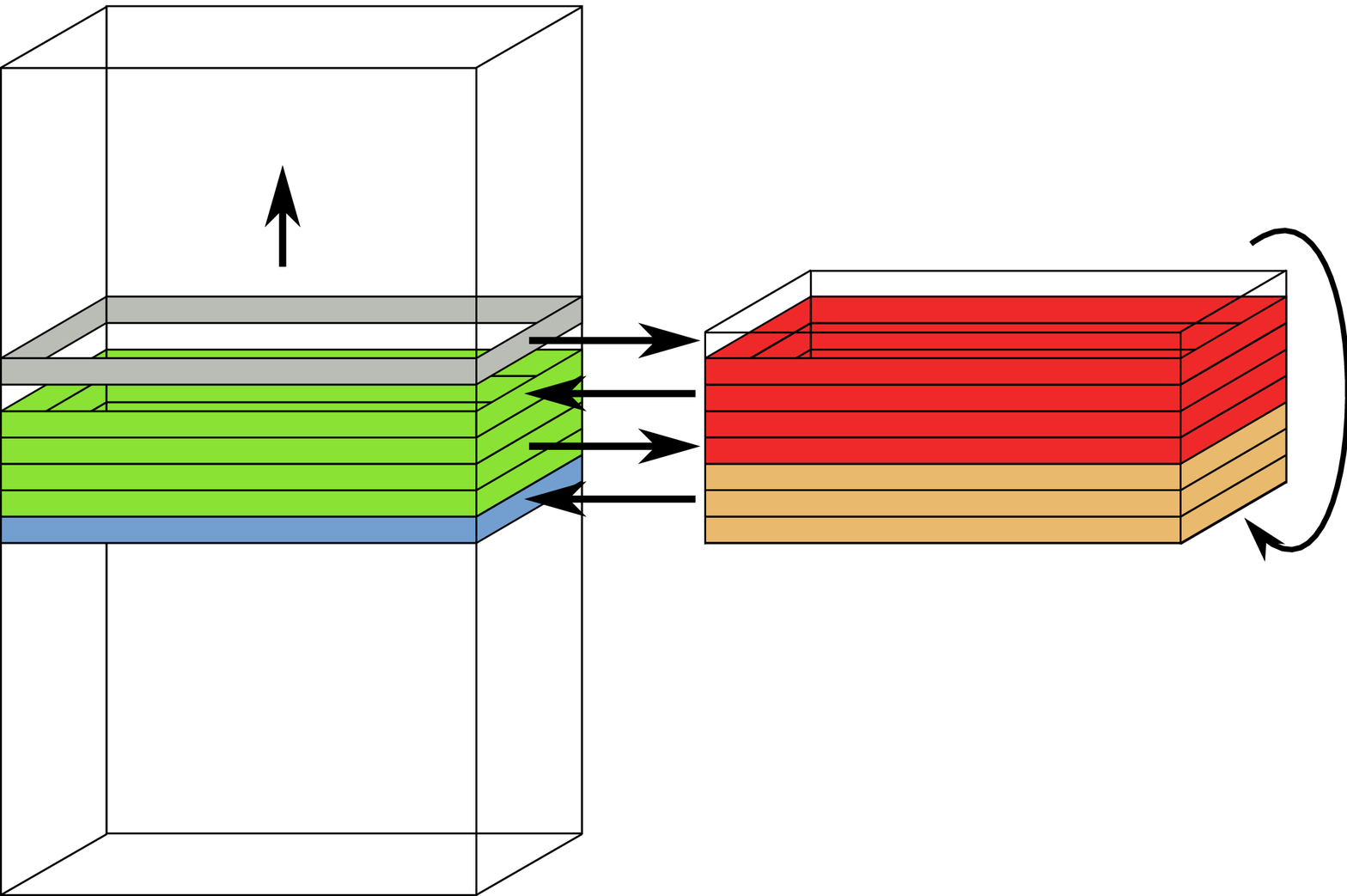}}
\caption{Temporal wavefront blocking (1 thread group with four threads)}
\label{fig:temporalBlocking}
\end{figure}
\begin{figure}[htbp]\centering
    \includegraphics*[width=0.9\linewidth]{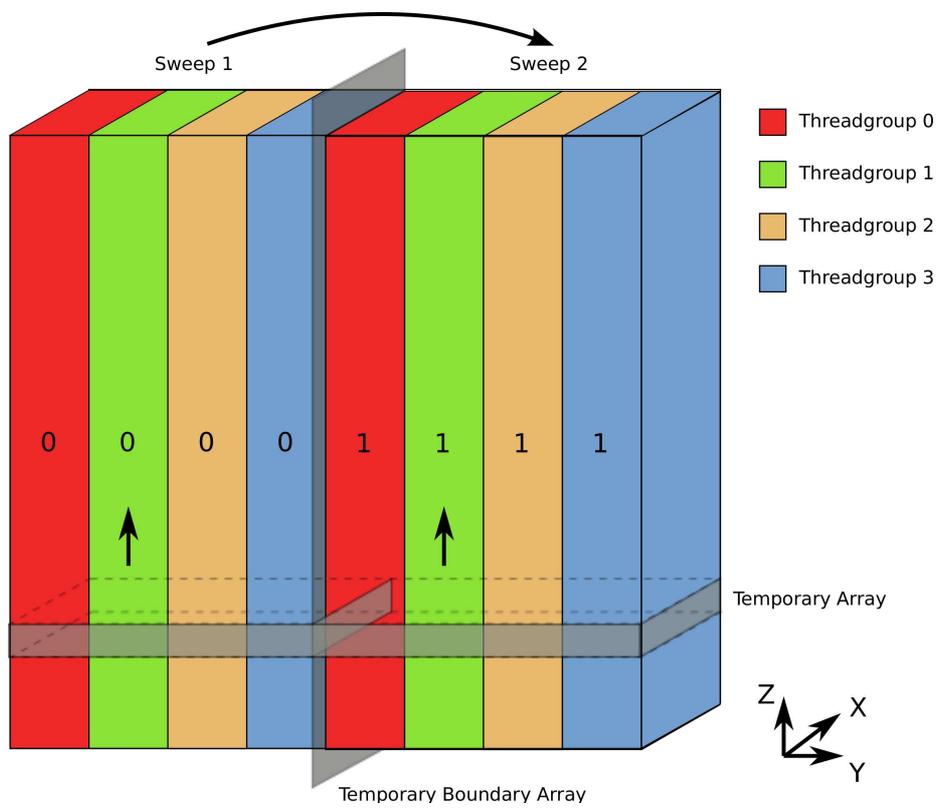}
    \caption{Organization of thread groups with spatial blocking}
    \label{fig:spatial}
\end{figure}
\begin{figure}[htbp]
    \centering
	\includegraphics*[width=1.0\linewidth]{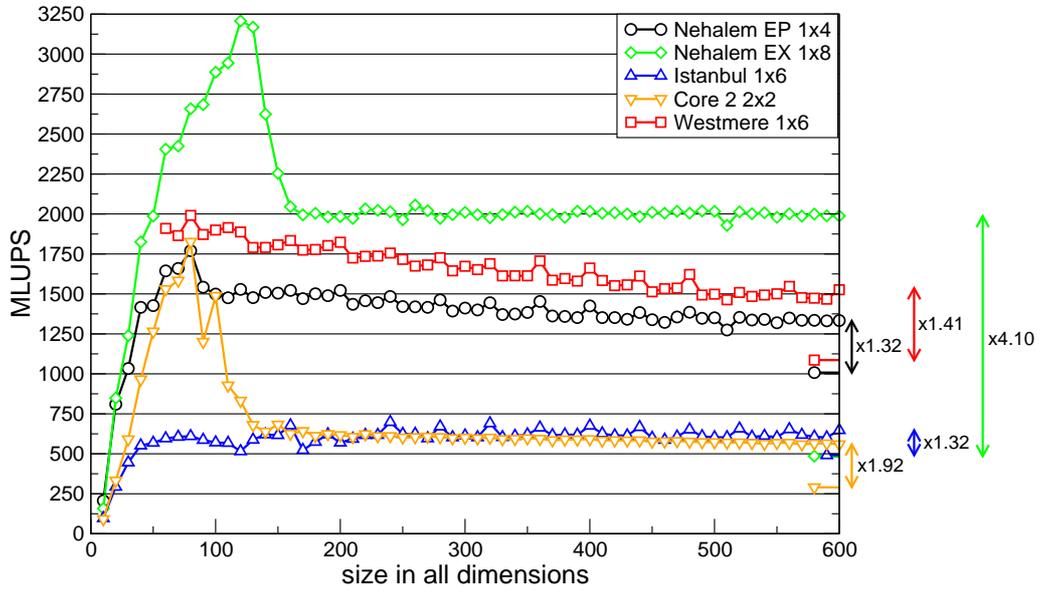}
    \caption{Wavefront temporal blocking results for the Jacobi smoother}
    \label{fig:wave_jacobi}
\end{figure}
%
%
\begin{figure}[htbp]
    \centering
	\includegraphics*[width=1.0\linewidth]{wavefront-gs-all.eps}
    \caption{Wavefront temporal blocking results for the Gauss-Seidel smoother}
    \label{fig:wave_gauss_seidel}
\end{figure}
\begin{figure}[htbp]
    \centering
	\includegraphics*[width=1.0\linewidth]{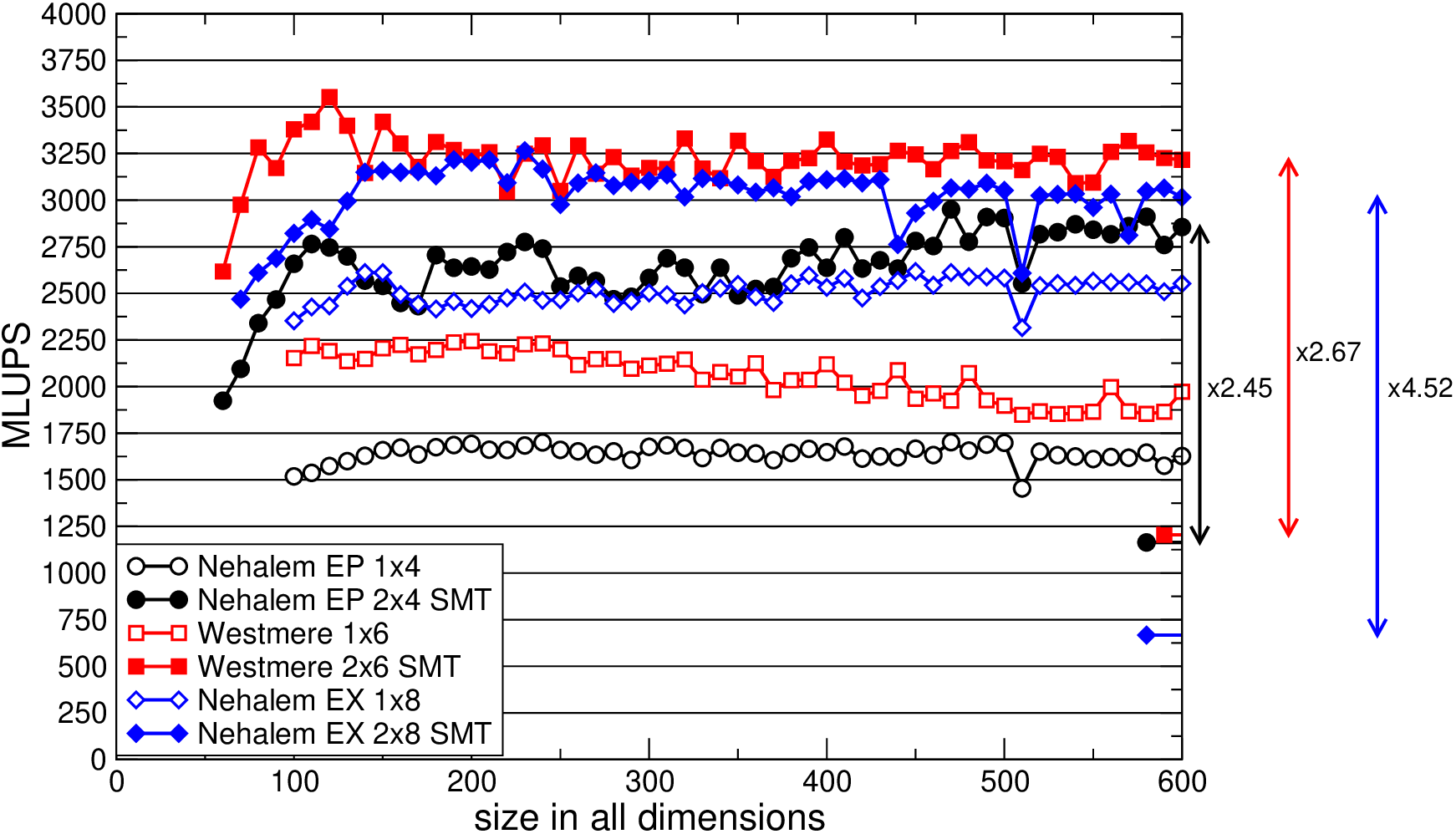}
    \caption{Wavefront temporal blocking results for the Gauss-Seidel smoother with SMT}
    \label{fig:wave_gauss_seidel_smt}
\end{figure}

Our wavefront parallelization technique implements implicit temporal blocking
by utilizing the property of modern multicore architectures that multiple cores
share the outermost cache level. The grid is decomposed into blocks. A block is
updated by a ``thread group,'' consisting of a number of threads. Each thread
in a thread group performs one sweep on the block, successively updating
the ``planes'' in $z$ direction. Planes updated by
consecutive threads are guaranteed to be still located in the shared cache.
This update mechanism is illustrated in Fig.~\ref{fig:temporalBlocking}.
Because multiple updates are performed while holding the data in cache and the
intermediate update steps need not be stored, the second grid of the
out-of-place Jacobi method is not required.  Odd-numbered updates are instead written
to a temporary array, which is large enough to hold the intermediate steps of
all updates until the last update is written back to the \verb.src. array. In step
(a) the first thread of a group performs the first plane update,
loading from \verb.src. and writing to a small temporary array. With a distance of two to
ensure the correct update order, the second thread performs
the second update from the temporary array back to the \verb.src. array in step 
(b)\@. In our
example with four threads another two updates are performed until the fourth
update is written to the \verb.src. array. The temporary array is used in a round
robin manner, and hence shifted through the $z$ dimension of the spatial grid. It must be large
enough to hold the needed \verb+dst+ planes of all threads, for our example eigth.
The threads have to be synchronized after each plane update, and block sizes must
be chosen so that the temporary data can be kept in the
outermost cache level. Since the number of thread groups, the number of
threads per thread group and the block size can be freely configured this
scheme can map to the underlying hardware in a very flexible way. A drawback of
the current implementation is that the maximum number of blocked updates is
determined by the number of threads available. This method requires fine-grained 
parallelism, making it crucial to employ an efficient synchronization
mechanism. Our threaded implementation is based on the POSIX thread API. The
pthread barrier turned out to have a very large overhead, making it unsuitable
for fine-grained parallelism.  For small thread counts as applicable on a single
socket, an implementation of a spin waiting loop was used for the barrier.
Since this does not perform well with SMT threads, a tree barrier was implemented which provided
less overhead whenever more than one logical thread per core was used.

The spatial blocking scheme is illustrated in Fig.~\ref{fig:spatial}.
Every thread group performs a parallel wavefront update as explained above.
The domain is decomposed into $B$ blocks along the $y$ dimension (8 in \ref{fig:spatial}). Each thread
group works on one or more blocks. All thread groups update each block in a
synchronized fashion, and one $z$ sweep is performed on the first $N$ blocks,
where $N$ is the number of thread groups.  
The boundary between consecutive sweeps must be set up so that
the next sweep can proceed with correct boundary conditions on the interface
to the previous sweep. If $t$ is the number of threads in a group,
a boundary array must thus hold $t$ planes in $z$-$x$ direction.
Hence no additional computations are necessary for the boundary treatment.
%

Figure~\ref{fig:wave_jacobi} shows the results for Jacobi with temporal wavefront
blocking on one socket. The baselines drawn on the right are the
threaded results without temporal blocking for a problem size of 200x200x200.
On Core 2 there is a very large gap between in-cache and in-memory memory baselines, 
indicating a
considerable potential for temporal blocking. A speedup of two could be achieved
by wavefront temporal blocking, but to
leverage the potential on this architecture a bigger blocking factor (i.e., more
cores per thread group) would be
required in order to decouple from main memory bandwidth.  
A completely different picture can be seen on the Nehalem EP. Here
the gap between in-cache and memory performance is rather small. The threaded
memory performance utilizing non-temporal stores is already 1008~MLUPS. 
While a speedup of 25--50\,\%  seems fair, the scaling of 
memory bandwidth with the number of cores results in a high baseline
and limits the benefit of our optimization. The result on the Westmere
processor is similar, still there is a speedup benefit from the higher blocking
factor of six. The Intel Nehalem EX in the configuration used here shows the highest
benefit. It allows a blocking factor of
eight together with a very high bandwidth L3 cache. On the other hand,
this (artificial) configuration is strongly bandwidth-starved in terms
of available bandwidth per core. This combination results in speedups
of four independent of the problem size. Istanbul has a good initial
position promising a significant speedup, there is a large gap between
saturated in-cache and main memory performance, and the number of cores per
socket allows a blocking factor of six. However, it only achieves speedups 
comparable to Nehalem EP.

Our scheme for temporal wavefront parallelization can be adapted for the
in-place Gauss-Seidel method if used in combination with the pipeline parallel
approach. Since all updates operate on one array, additional temporary arrays
are not required. To ensure the correct update ordering the updates have to be
shifted for lexicographic Gauss-Seidel between thread groups as illustrated in
Fig.~\ref{fig:pipeline_gs} (b). This is a natural extension to the threaded
pipelined parallelization introduced in Sec.~\ref{sec:code_analysis}.
The results for temporal wavefront blocking with Gauss-Seidel are shown in
Fig.~\ref{fig:wave_gauss_seidel}. Again the baseline of a threaded Gauss-Seidel
with pipeline parallelization for a size of 200x200x200 is indicated on the right
$y$ axis.  On the Core 2 the combination of low (and non-scaling) main memory
performance together with only a blocking factor of two leads to
a speedup of nearly two for temporal blocking. Nehalem EP shows
improvements of 30--40\,\%. Westmere profits from its two additional
cores and hence from the deeper blocking factor, reaching a speedup
of over 50\,\%. The potential of our optimization technique is again seen on the
example of Nehalem EX. Despite its hardware configuration yielding
the lowest bandwidth compared to the other Nehalem processors, it
reaches the best overall performance and can fully benefit from its eight cores
and strong L3 cache subsystem, showing an impressive speedup of 3.8. 
The Istanbul architecture again shows disappointing results comparable 
to the Nehalem EP. Its exclusive cache hierarchy seems to be unsuited for these
bandwidth-demanding in-cache loops, but the exact reason for the small
gains of our optimization were not investigated in more detail.

As noted in Sec. \ref{sec:code_analysis}, Gauss-Seidel cannot be fully
pipelined due to its recursive structure. While our
optimized kernel reduces this penalty, it is still noticeable, causing
the floating point units to be underutilized. For exactly this
situation, a shared resource being not fully used, the Nehalem
processors implement SMT with two hardware threads sharing a
physical core. Further possible benefits of using SMT threads for our
temporal wavefront optimization are manifold, deeper blocking
factors being just one effect. Since main memory bandwidth on the Nehalem
(EP and Westmere) processors scales with the number of threads, the utilization of
the memory bus can be increased by using multiple thread groups. 
And finally, two SMT threads also share the L1 and L2 caches, potentially
reducing necessary cacheline transfers in our pipelined wavefront
setting.  Figure~\ref{fig:wave_gauss_seidel_smt} shows the results,
with filled symbols denoting SMT data. Nehalem EP and Westmere now achieve
speedups of 2.5 versus the threaded baseline.  The SMT benefit on
Nehalem EX is not that large, which could be caused by the fact that
this chip was already arithmetically limited. An indication
supporting this conjecture is that Nehalem EP Westmere and Nehalem
EX now reach a comparable performance in accordance to their similar
arithmetic peak performance per socket. Nehalem EX can reach an overall
speedup of up to five against its threaded baseline.
%
\section{Conclusion}
\label{sec:summary}
%
We have presented a novel way to implement temporal blocking,
specifically designed to leverage the shared outer-level cache on 
today's multicore architectures. The
optimizations were evaluated on a wide range of current multicore processors to show their
potential. Besides the well-known Jacobi method we have presented a highly
efficient implementation of the recursive Gauss-Seidel method. For the threaded
and temporal wavefront implementation of Gauss-Seidel we employed a pipeline
parallel approach, retaining the update ordering of the serial algorithm. The
optimization reaches considerable speedups on all architectures. Results
on our (artificially) bandwidth-starved Nehalem EX system confirm
that a large ratio between in-cache and memory bandwidths improves the
gain for our temporal blocking approach. The cache subsystem of the AMD
Istanbul turned out to be incapable of benefitting from our optimizations to the
same extent as comparable Intel architectures.  Finally we have
shown that employing SMT threads for temporal blocking of the Gauss-Seidel
solver yields large performance improvements with overall speedups of up to five 
on Intel Nehalem EX. It is noteworthy that for this case both Nehalem EP 
and Nehalem EX reach their full arithmetic potential independent of their 
different memory bandwidth capabilities.
\section*{Acknowledgment}
We are indebted to Intel Germany for providing test systems and early access
hardware for benchmarking.


\small\begin{thebibliography}{XX}
\bibitem{bergen05} B.~Bergen, F.~H\"ulsemann, U.~R\"ude: \emph{Is
  1.7$\times$10$^{10}$ Unknowns the Largest Finite Element System that
  Can Be Solved Today?} In: ACM/IEEE (Ed.): Proceedings of the
  ACM/IEEE SC 2005 Conference (Supercomputing Conference '05, Seattle,
  Nov 12--18, 2005).

\bibitem{hager03} G.~Hager, F.~Deserno, G.~Wellein: \emph{Pseudo-Vectorization and RISC Optimization                                                             
Techniques for the Hitachi SR8000 Architecture}
    In: High Performance Computing in Science and Engineering Munich 2002 (editor S. Wagner et al.), 
    Springer, pp.\ 425-442, 2003

\bibitem{datta08} K.~Datta, M.~Murphy, V.~Volkov, S.~Williams,
  J.~Carter, L.~Oliker, D.~Patterson, J.~Shalf, K.~Yelick:
  \emph{Stencil Computation Optimization and Auto-tuning on
    State-of-the-Art Muticore Architectures.} In: ACM/IEEE (Ed.):
  Proceedings of the ACM/IEEE SC 2008 Conference
  (Supercomputing Conference '08, Austin, TX, Nov 15--21, 2008).
 
\bibitem{datta09} K.~Datta, S.~Kamil,  S.~Williams, L.~Oliker, J.~Shalf, K.~Yelick:
  \emph{Optimization and Performance Modeling of Stencil
  Computations on Modern Microprocessors.} In: SIAM Rev.:
  vol. 51, No.\ 1, pp.\ 129--159,  2009.

\bibitem{wellein09} G. Wellein, G. Hager, T. Zeiser, M. Wittmann and H. Fehske: 
  \emph{Efficient temporal blocking for stencil computations 
    by multicore-aware wavefront parallelization.} 
  Proc. COMPSAC 2009.  DOI:\texttt{10.1109/COMPSAC.2009.82}

\bibitem{kowarschik04} M.~Kowarschik: \emph{Data Locality
    Optimizations for Iterative Numerical Algorithms and Cellular
    Automata on Hierarchical Memory Architectures.} 
  PhD thesis, July 2004, SCS Publishing House, Germany. ISBN 3-936150-39-7.

\bibitem{treibig09} J.~Treibig: \emph{Efficiency Improvements of Iterative Numerical Algorithms on Modern Architectures.} 
  PhD thesis, July 2009, URN: urn:nbn:de:bvb:29-opus-14036.

\bibitem{frigo99} M.~Frigo, C.\,E.~Leiserson, H.~Prokop,
  S.~Ramachandran: \emph{Cache-Oblivious Algorithms.} In: 40th Annual
  Symposium on Foundations of Computer Science, FOCS 99, Oct 17--18, 1999, 
  New York, NY\@.
  
\bibitem{zeiser08} T.~Zeiser, G.~Wellein, A.~Nitsure, K.~Iglberger,
  U.~R\"ude, G.~Hager: \emph{Introducing a parallel cache oblivious
    blocking approach for the lattice Boltzmann method.} 
  Progress in CFD, vol. 8, No.\ 1--4, pp.\ 179--188,  2008.

\bibitem{kerby06} D.\,J.~Kerbyson, A.~Hoisie: \emph{Analysis of
    Wavefront Algorithms on Large-scale Two-level Heterogeneous
    Processing Systems.} In Proc. Workshop on Unique Chips and Systems
    (UCAS2), IEEE Int. Symposium on Performance Analysis of Systems
    and Software (ISPASS), Austin, TX, 2006.  

\bibitem{stream}J.D.~McCalpin: \emph{STREAM: Sustainable Memory 
    Bandwidth in High Performance Computers.} 
  http://www.cs.virginia.edu/stream
 
\bibitem{likwid} \emph{LIKWID tool suite.} 
    http://code.google.com/p/likwid/

\bibitem{georg10} M.~Wittmann, G.~Hager, G.~Wellein: \emph{Multicore-aware
    parallel temporal blocking of stencil codes for shared and distributed
    memory}. Accepted for LSPP10, the Workshop on Large-Scale Parallel
    Processing at IPDPS 2010, April 23rd, 2010, Atlanta, GA. arXiv:0912.4506

\bibitem{Jan09} J. Treibig and G. Hager: \emph{Introducing a
      Performance Model for Bandwidth Limited Loop Kernels}.
     Workshop on Memory Issues on Multi- and Manycore Platforms,
    PPAM2009.
\end{thebibliography}
\end{document}